# Introducing nanoengineering and nanotechnology to the first year students through an interactive seminar course

Hassan Raza[1], Tehseen Z. Raza[2]

[1] Department of Electrical and Computer Engineering, University of Iowa, Iowa City, Iowa 52242, USA
nstnrg@gmail.com
[2] Department of Physics and Astronomy, University of Iowa, Iowa City, Iowa 52242, USA
tehseen-raza@uiowa.edu

Abstract: We report a first year seminar course on nanoengineering, which provides a unique opportunity to get exposed to the bottom-up approach and novel nanotechnology applications in an informal small class setting early in the undergraduate engineering education. Our objective is not only to introduce the fundamentals and applications of nanoengineering but also the issues related to ethics, environmental and societal impact of nanotechnology used in engineering. To make the course more interactive, laboratory tours for microfabrication facility, microscopy facility, and nanoscale laboratory at the University of Iowa are included, which inculcate the practical feel of the technology. The course also involves active student participation through weekly student presentations, highlighting topics of interest to this field, with the incentive of "nano is everywhere!" Final term papers submitted by the students involve a rigorous technology analysis through various perspectives. Furthermore, a student based peer review process is developed which helps them to improve technical writing skills, as well as address the ethical issues of academic honesty while reviewing and getting introduced to a new aspect of the area presented by their colleagues. Based on the chosen paper topics and student ratings, the student seemed motivated to learn about the novel area introduced to them through theoretical, computational and experimental aspects of the bottom-up approach.

## 1. INTRODUCTION

Nanoengineering is an emerging field, which is leading to novel nanotechnology [1] applications due to atomic scale engineering. On the other hand, it is a new paradigm in fundamental thinking and understanding about the physical universe, where the bottom up approach is the norm and not an exception. In this new approach, one has to think in terms of atoms and how they interact to make useful materials, structures, devices and systems [2-4]. Being on the cross-roads of this paradigm shift in both research and education [5-7], it becomes imperative to introduce concepts related to nanoengineering at an early stage to the engineering undergraduate students.

There have been numerous efforts to introduce nanotechnology in the engineering and science curriculum [8-20]. In this paper, we report a first-year seminar course focused on introduction to nanoengineering and its ethical and societal impact. The course comprises of a series of informal "discussion-style" lectures to introduce students to the "bottom-up" manufacturing at the atomic, molecular and nano scales. To ensure the informal communication, class size was restricted to fifteen students. Furthermore, these lectures educate the students about the complete engineering design process at the nanoscale starting with the nanoscale materials and structures, to nanoscale devices and systems. Additionally, the convergence of various engineering and science disciplines at the nanoscale [5] is taught, which makes nanoengineering and nanotechnology truly multi-disciplinary. The goal of this course is to convey the breadth, depth and diversity of the topics required to give a big, yet comprehensive picture of nanoengineering for potential applications in electronics, spintronics, solar cells, thermoelectric devices, batteries, fuel-cells, nano-electromechanical systems (NEMS), physical, chemical and bio-sensors, while also being vigilant of various societal, ethical [21,22,23], global [24] and environmental [25,26] challenges of the potential use of this technology.

To achieve these objectives, the reported course covers the following aspects:

- **Bottom-Up approach:** The set of lectures start with developing an atomistic approach towards the materials and structures, then understanding various devices based on the material characteristics and finally more complex systems comprising of various devices. This introduces them to the bottom-up approach for this technology, as opposed to the top-down approach of the conventional engineering design.

- **Theory, computation and experiments:** Most of the science, engineering and technology now is divided into the above stated three disciplines. In this course, we try to give a big picture about how theory, computation and experiments work together to engineer devices at the nanoscale.

- **Laboratory Tours:** Various laboratory tours to the microfabrication facility, microscopy facility and nanoscale laboratory at the University of Iowa were planned to give students a practical feel of how devices are fabricated, characterized and integrated.

- **Silicon Run Documentary:** As a part of introducing the industrial infrastructure to this area, Silicon Run documentary part I and part II [27] were shown in class, which is followed by interactive discussion about complement fabrication and characterization. Although these documentaries address top-down approach, students found them very useful in the context of nanoscale fabrication and characterization.

- **Student participation:** Another key aspect is active student participation by providing a seminar and a term-paper opportunity to them. This not only motivates students to study the state-of-the-art fundamental and applied research but also makes them conscious of the complete analysis of nanotechnology design space. As a part of the course, students are required to prepare a 15 minute presentation on the topic of their interest related to nanoengineering and nanotechnology. Students also write a term paper, which is peer-reviewed by their colleagues in class to acquaint both the authors and the referees with the necessary aspect of academic honesty, ethics and reviewing others scientific and engineering work [28]. Incorporating the peer review process, in otherwise standard term-paper assignment, gives students an opportunity to get prepared for the peer-review process of academics and now various industrial organizations. It also gives them opportunity to review and analyze a totally different topic presented by their colleague. The presentation and the term-paper opportunity also motivated students to explore the area in the direction of their interest and in the quest of understanding novel concepts; they learnt to apply them to nanoengineering.

Additionally, the objective of the course is to engage students in scientific inquiry and engineering design process to retain them in engineering disciplines. The instructional strategies promoted the criteria recommended by National Research Council. The small class sized helped the group to become a tight-knit community of engineers discussing scientific, industrial, economic aspects of this new technology and also its impact on environment and society. They not only developed a sense of belonging to the group, but it also helped them get comfortable with the academic environment that is important for a first year student. The nature of contributions in terms of presentations and peer review process imparted sense of responsibility and pride by observing and maintaining high ethical standards to avoid plagiarism. These ethical standards need to be a part of basic introductory exposure to any area, of which we try to emphasize in this course.

## 2. TEACHING APPROACH

This seminar course is intended to excite the incoming students to the emerging field of nanoengineering to produce the next generation of scientists and engineers. The course introduces students to the basic concepts, techniques and tools that are central to the rapidly developing field of nanoengineering and to some of the important scientific and technological innovations that are now emerging or are expected to emerge, addressing their ethical, societal and environmental impacts. An expected outcome of the course is that the students should be able to realize nanoengineering as a new paradigm in fundamental thinking and to take advantage of the atomic and nanoscale opportunities to develop novel nanotechnology applications. To achieve these objectives, we discuss the following topics in an introductory manner in a fairly informal setting:

- Introduction to nanoengineering (to educate how engineering at the nanoscale is different and how it offers novel opportunities for nanotechnology applications)

- Particle, waves and duality (to convey the dual nature of matter waves)

- Atomic matter (to teach about the matter waves)

- Atomic structure (how arrangements of atoms determine their physical properties)

- Electronic structure (how electronic structure can be understood connecting with the atomic structure)

- Quantum transport (how current flows at the nanoscale)

- Nanoscale devices (how devices work at the nanoscale for various applications)

- Nanoscale systems (how devices can be connected together to form systems)

- Nanofabrication (how devices and systems are made; introduction to various lithography and fabrication techniques, aided by Silicon Run documentaries [27] and various laboratory tours).

- Characterization at the nanoscale (how materials, devices and systems can be characterized; introduction to various spectroscopic and microscopic techniques)

Furthermore, like any new technology, nanotechnology also raises many ethical and societal issues, including concerns about the toxicity and the environmental impact of the nanomaterials and devices. In this course, we bring these concerns to light and discuss them openly to create awareness amongst the students and make them conscious of their responsibilities as citizen engineers. With every new technology, addressing safe practices should always be emphasized. The reported seminar course incorporates such awareness seminars regularly. Students are encouraged to address one of the topics relating to the societal impact of nanoengineering and nanotechnology in their presentations. The following topics were given as examples of seminar topics to the students:

- Sustainability in Nanotechnology
- Global issues
- Ethics in nanotechnology
- Safety in nanotechnology
- Green nanotechnology manufacturing
- Green energy harvesting

## 3. STUDENT SEMINARS AND TERM PAPERS:

The diverse choice of presentations and term papers available online [29] by students shows a broad impact on their scientific perspective about nanoengineering and nanotechnology inculcated through this course. For convenience, the topics chosen by the students are divided into the following broad categories,

*Manufacturing and Economy*:

Safety concerns of nanotechnology

Nanotechnology and its influence on economy

Nanotechnology in building construction

Nanotechnology improving our transportation in the near future

Bottom-up production and Nanobots

Molecular nanotechnology

*Industrial and Consumer Applications*:

Carbon nanotubes

The marvels of nanotechnology and its applications in telecommunications

Everyday applications of nanotechnology

*Computing Applications*:

Nanotechnology in computers

How nanotechnology has shaped todays computers

Quantum computing

*Food related Applications*:

Nanotechnology in food processing

Nanotechnology in food storage

Nanotechnology and water treatment

*Medical Applications*:

Nanotechnology in drug delivery

Nanotechnology in cancer treatment

Nanotechnology in therapeutics and imaging

*Military Applications*:

Military Implications of nanotechnology

Nanoscience in the future soldier

Science fiction or science fact: Nanotechnology in defense

## 4. LABORATORY TOURS AND DEMONSTRATIONS

As a part of the course, the students visit the "Micro Fabrication Facility", the "Microscopy Facility" and the "Nanoscale Devices and Systems Laboratory" at the University of Iowa, to give them exposure to the following laboratory equipment (about which they also study in the class).

- Various lithography equipment including mask aligner, photolithography, electron-beam/thermal evaporators, reactive ion etchers, atomic layer deposition system, etc; and characterization equipment including optical profiler, mechanical profiler, ellipsometer, etc during the Micro Fabrication Facility visit.

- Various microscopy equipment, including SEM (scanning electron microscope), AFM (atomic force microscope), TEM (transmission electron microscope), Confocal Raman, etc, during the Microscopy Facility visit.

- CVD (chemical vapor deposition) furnace and various other graphene and CNT (carbon nanotube) sample preparation equipment during the Nanoscale Devices and Systems Laboratory visit.

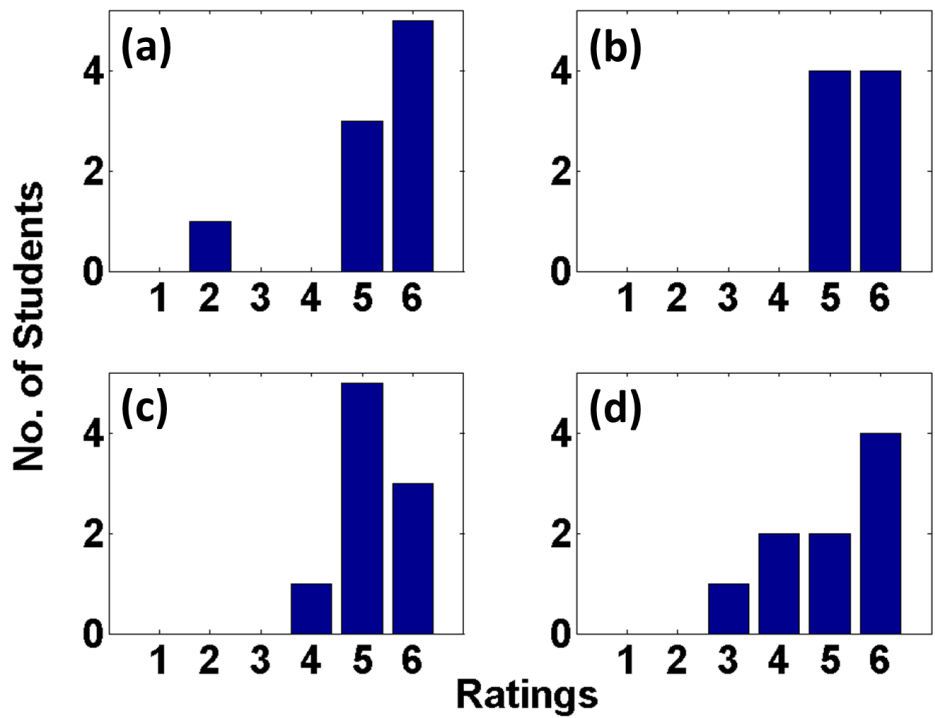

Figure 1: Student feedback distribution. (a) Effectiveness of Instruction, (b) Acquiring a basic understanding of nanoengineering, (c) Organization of the course, (d) Contribution of the instructional methods to learning.

## 5. STUDENT SURVEY RATINGS:

This course was successfully offered in Fall 2010, with a repeat in Fall 2011 in the College of Engineering at the University of Iowa as an elective course for the incoming first year students. Twenty seven students participated in the course, out of which three students were female and one student belonged to African-American minority group. Apart from freshman standing, there was no pre-requisite for the course.

Based on the student feedback, the course enhanced their interest in the nanoscale engineering of nanotechnology and equipped them well for further academic and/or industrial pursuit in this field. The feedback comments are as below, followed by Table I and Fig. 1 giving the statistical comparison for various aspects of the course.

- It covered a lot of information over a short amount of time fairly efficiently.

- I learned not only applications of nanotechnology but how it works.

- The course was very relaxed and I enjoyed going.

- This seminar was fascinating and a great introduction to nano-science. I really enjoyed the discussions each week, and I liked working on my presentation.

- The course did a good job of introducing the topics of nanotechnology and providing a good basis of quantum chemistry.

Table I: Statistics of student feedback.

|  | Fall 2010 | | Fall 2011 | |
| --- | --- | --- | --- | --- |
|  | Mean | Median | Mean | Median |
| Effectiveness of Instruction | 5.75 | 6.00 | 4.80 | 5.00 |
| Acquiring a basic understanding of nanoengineering | 5.75 | 6.00 | 4.40 | 5.00 |
| Organization of the course | 5.5 | 5.50 | 5.00 | 5.00 |
| Contribution of the instructional methods to learning | 5.25 | 5.50 | 4.80 | 5.00 |

Ratings:
6 = Strongly Agree
5 = Moderately Agree
4 = Slightly Agree
3 = Slightly Disagree
2 = Moderately Disagree
1 = Strongly Disagree

While it was not feasible to associate the satisfaction level to various aspects of the course due to a generic format used by the College of Engineering, it was reassuring that the students rated the course quite well. Additionally, in Fall 2011, the student satisfaction dropped compared to Fall 2010 – the reasons for which are still unclear. But overall, in both the offerings, the student participation was active. We propose a few future developments to further enhance the experience, where an effort also will be made to encourage students to share their newly researched information with public and greater community.

## 6. FUTURE DEVELOPMENTS:

**6.1 Atomic visualizations.** Motivated by the bottom-up approach, we plan to use atomic visualization tools [30,31] to complement lecture teaching about zero-dimensional materials (e.g. molecules, quantum dots, nanocrystals, etc.), one-dimensional materials (e.g. carbon nanotubes [32], silicon nanowires, graphene nanoribbons [33], etc.), two-dimensional materials (e.g. graphene [33]) and finally three-dimensional materials (e.g. magnetic tunnel junctions, heterostructures, etc) of relevance to nanotechnology. In short, the objective is to incorporate the reported work [34-47] in the educational component [48].

**6.2 Microscopy and spectroscopy demonstration.** We plan to have a demonstration of imaging carbon nanotubes using AFM and SEM. Furthermore, we plan to give student a demonstration of Raman spectroscopy of CNTs to measure the diameter by analyzing the radial breathing modes.

**6.3 Fabrication and characterization demonstration.** We propose to give a demonstration of photolithography and electrical characterization by fabricating Flash memory devices using nanomaterials, like Fullerenes.

**6.4 Projects to "make something useful Nano under $5".** The course would have a project at the end of semester where students will define a useful nanodevice (natural or man-made) under $5. They would discuss various technological and societal impact of this device in the final presentation. The examples may include inexpensive laser pointers, Flash memories, creams with nanoparticles, etc.

**6.5 Nanoday:** As a part of the final course project, students will be encouraged to display their work in a poster presentation. Such an activity will not only advertise the students work and excite newcomers to this area, but also give them an opportunity of conference atmosphere and confidence of answering questions on one-to-one basis.

## 7. CONCLUSIONS:

We have successfully developed a first-year seminar course on nanoengineering and nanotechnology. This elective course has been incorporated in the regular engineering curriculum at the University of Iowa. The students will be equipped with the theoretical knowledge base and the right skills to propose, fabricate and then analyze the nanoscale devices. The critical skills will not only help them evaluate the technology merits but also the overall impact on the society and our environment. The broader aspects of the reported seminar course will help produce engineers with the right skills to solve the grand challenges that our civilization faces with the help of nanoengineering and nanotechnology. This educational endeavor will train high-tech well-trained environmentally conscious engineers in nanotechnology, which will be ready to solve the technological and societal problems of today and tomorrow. Further developments of the reported seminar course are also discussed.

**Acknowledgements:**

We would like to acknowledge the Provost office at the University of Iowa for financial support.